%% file: main.tex
\documentclass[sigconf, nonacm]{acmart}
\AtBeginDocument{%
  }

\usepackage{graphicx}
\usepackage{booktabs}
\usepackage{amsmath}
\usepackage{placeins}
\usepackage{float}
\usepackage{enumitem}
\usepackage{tabularx}
\usepackage{array}
\usepackage{threeparttable}

\AtBeginDocument{%
  \setlength{\abovedisplayskip}{1pt}
  \setlength{\belowdisplayskip}{1pt}
  \setlength{\abovedisplayshortskip}{1pt}
  \setlength{\belowdisplayshortskip}{1pt}
}

\setcopyright{acmlicensed}
\copyrightyear{2018}
\acmYear{2018}
\acmDOI{XXXXXXX.XXXXXXX}
\acmConference[Conference acronym 'XX]{Make sure to enter the correct
  conference title from your rights confirmation email}{June 03--05,
  2018}{Woodstock, NY}
\acmISBN{978-1-4503-XXXX-X/2018/06}

\begin{document}

\title[HEMERA: A Memory-Centric Accelerator for SSD Models]
{HEMERA: A Heterogeneous Memory-Centric Accelerator with
Recursive Dataflow for Edge-Constrained
State-Space-Duality Models Inference}


\author{Hao Ding}
\affiliation{%
  \department{School of Integrated Circuits}
  \institution{Peking University}
  \city{Beijing}
  \country{China}
}
\affiliation{%
  \institution{Unigroup Guoxin Microelectronics Co., Ltd.}
  \city{<official city>}
  \country{China}
}

\author{Ling Liang}
\email{Lingliang@pku.edu.cn}
\affiliation{%
  \department{School of Integrated Circuits}
  \institution{Peking University}
  \city{Beijing}
  \country{China}
}

\author{Ruitong Qiao}
\affiliation{%
  \department{School of Integrated Circuits}
  \institution{Peking University}
  \city{Beijing}
  \country{China}
}

\author{Xiantong Qiu}
\affiliation{%
  \department{School of Integrated Circuits}
  \institution{Peking University}
  \city{Beijing}
  \country{China}
}

\author{Dongxue Zhao}
\affiliation{%
  \department{School of Integrated Circuits}
  \institution{Peking University}
  \city{Beijing}
  \country{China}
}

\author{Jinshan Li}
\affiliation{%
  \department{School of Integrated Circuits}
  \institution{Peking University}
  \city{Beijing}
  \country{China}
}

\author{Lei Jin}
\affiliation{%
  \institution{Yangtze Memory Technologies Co., Ltd.}
  \city{Wuhan}
  \country{China}
}

\author{Zhiliang Xia}
\affiliation{%
  \institution{Yangtze Memory Technologies Co., Ltd.}
  \city{Wuhan}
  \country{China}
}

\author{Zongliang Huo}
\affiliation{%
  \institution{Yangtze Memory Technologies Co., Ltd.}
  \city{Wuhan}
  \country{China}
}

\author{Meng Li}
\affiliation{%
  \department{School of Integrated Circuits}
  \institution{Peking University}
  \city{Beijing}
  \country{China}
}

\author{Zongwei Wang}
\email{wangzongwei@pku.edu.cn}
\affiliation{%
  \department{School of Integrated Circuits}
  \institution{Peking University}
  \city{Beijing}
  \country{China}
}
\affiliation{%
  \institution{Beijing Advanced Innovation Center for Integrated Circuits}
  \city{Beijing}
  \country{China}
}
\affiliation{%
  \institution{YanXin MicroElectronics Co., Ltd. (YXME)}
  \city{<official city>}
  \country{China}
}

\author{Yimao Cai}
\email{caiyimao@pku.edu.cn}
\affiliation{%
  \department{School of Integrated Circuits}
  \institution{Peking University}
  \city{Beijing}
  \country{China}
}
\affiliation{%
  \institution{Beijing Advanced Innovation Center for Integrated Circuits}
  \city{Beijing}
  \country{China}
}

\renewcommand{\shortauthors}{Ding et al.}

\makeatletter
\renewcommand{\@mkauthors}{%
  \begingroup
  \hsize=\textwidth
  \global\setbox\mktitle@bx=\vbox{%
    \noindent\unvbox\mktitle@bx
    \par\medskip
    \centering

    \begin{minipage}{0.98\textwidth}
      \centering

      {\normalsize
      Hao Ding\textsuperscript{1,4},
      Ling Liang\textsuperscript{1,\ensuremath{\dagger}},
      Ruitong Qiao\textsuperscript{1},
      Dongxue Zhao\textsuperscript{1},
      Xiantong Qiu\textsuperscript{1},
      Jinshan Li\textsuperscript{1},
      Meng Li\textsuperscript{1},
      Lei Jin\textsuperscript{3},
      Zhiliang Xia\textsuperscript{3},\\[-1pt]
      Zongliang Huo\textsuperscript{3},
      Zongwei Wang\textsuperscript{1,2,\ensuremath{\dagger}},
      Yimao Cai\textsuperscript{1,\ensuremath{\dagger}}
      \par}

      \vspace{0.35em}

      {\normalsize
      \textsuperscript{1}Beijing Advanced Innovation Center
      for Integrated Circuits, School of Integrated Circuits,
      Peking University, Beijing, China\\[-1pt]
      \textsuperscript{2}YanXin MicroElectronics Co., Ltd. (YXME)\\[-1pt]
      \textsuperscript{3}Yangtze Memory Technologies Co., Ltd.\\[-1pt]
      \textsuperscript{4}Unigroup Guoxin Microelectronics Co., Ltd.
      \par}

      \vspace{0.25em}

      {\normalsize
      \textsuperscript{\ensuremath{\dagger}}Corresponding authors:
      Lingliang@pku.edu.cn, wangzongwei@pku.edu.cn, caiyimao@pku.edu.cn
      \par}

    \end{minipage}

    \par\medskip
  }%
  \endgroup
}
\makeatother

\renewcommand{\shortauthors}{Ding et al.}

\begin{abstract}
Structured State Space Models (SSMs), such as Mamba, enable efficient long-sequence modeling with linear time complexity. Recent implementations realize this capability through Structured State Space Duality (SSD), which transforms recursive state evolution into matrix-form computations. However, SSD introduces substantial system-level overheads, including quadratic intermediate materialization, irregular data movement, and prefix-dependent execution, leading to excessive memory traffic and bandwidth demand on conventional architectures. Although prior accelerators mitigate these overheads through optimized dataflows or compute-in-memory techniques, they largely retain matrix-oriented SSD execution and cannot simultaneously avoid quadratic intermediate storage and efficiently map dependency-bound state propagation.

This paper presents HEMERA, a heterogeneous memory-centric accelerator for efficient Mamba-2 inference. Rather than directly executing the matrix-form SSD computation, HEMERA reformulates it into an algebraically equivalent streaming-recursive dataflow that avoids quadratic intermediate storage while preserving the original computation. The resulting heterogeneous execution paradigm maps dense linear operations onto in-memory computing units and recursive state updates onto a dedicated streaming engine. Across Mamba-2 models ranging from 130M to 2.8B, HEMERA achieves average latency speedups of 1.4$\times$--3.6$\times$ and energy-efficiency improvements of 12.2$\times$--27.0$\times$ over the official optimized fused Mamba-2 kernel on NVIDIA A100. It further reduces the average SSD-related execution-time ratio across model scales to 14.12\% during long-sequence inference, demonstrating its potential for efficient deployment under edge constraints.
\end{abstract}




\maketitle

\section{Introduction}
The rapid growth of edge applications calls for efficient on-device Large Language Model (LLM) inference \cite{draft, easz, 10738209}. However, the widely used Transformer architecture suffers from quadratic complexity, leading to prohibitive memory and bandwidth demands on resource-constrained platforms \cite{lightmamba,11043793}. To address this, Mamba-2~\cite{mamba2} adopts a non-attentional paradigm based on Structured State Space Duality (SSD), achieving linear sequence scaling and strong potential for long-sequence inference \cite{10565926, tang2024vmrnnintegratingvisionmamba}. However, its practical deployment remains challenging due to SSD-induced data movement and computation overheads.

The challenges stem from two key characteristics of SSD. First, intra-chunk token interactions generate intermediate structures that grow quadratically with the chunk size, reaching up to 160~MB in large models (e.g., a 2.8B model with a chunk size of 256), far exceeding on-chip memory capacity~\cite{11042937,10.1145/3719664}. Second, heterogeneous operators with intrinsic recursive dependencies restrict parallel execution, resulting in bandwidth-bound performance dominated by data movement~\cite{10.1145/3725843.3756121,11159607,11044026}. As a result, SSD's theoretical scaling advantage is difficult to realize efficiently in practice.

Prior accelerators for SSD-based models, such as MambaOPU~\cite{11132895} and HLX~\cite{10.1145/3725843.3756115}, improve performance through operator fusion, pipelining, and locality optimization. However, they still follow matrix-oriented SSD execution and retain its dependency structure. Consequently, intermediate-data management and serialization overhead remain significant, especially for long sequences and large models.

Compute-in-memory (CIM) architectures reduce data movement by performing computation near memory, making them promising for bandwidth-intensive workloads~\cite{11043720}. However, SSD execution is not well aligned with existing CIM designs, as its frequent intermediate state generation and recursive updates introduce irregular dataflow and dynamic dependencies, deviating from the static and dense patterns typically assumed. Consequently, conventional CIM architectures cannot efficiently support the generation and update of intermediate states~\cite{10.1145/3676536.3676776}. Recent efforts such as MamCIMFlow~\cite{11311021} attempt to combine CIM-based projections with selective-state streaming. While they demonstrate the potential of CIM for SSM acceleration, they do not reformulate the full SSD contraction or jointly support heterogeneous SSD operators with scalable precision. As a result, there is a need to jointly address recursive serialization and operator heterogeneity.

To fully unlock CIM acceleration for end-to-end Mamba deployment, three fundamental challenges must be addressed. First, SSD generates a large volume of intermediate states from temporal interactions, leading to significant buffering overheads that overwhelm on-chip memory. Second, SSD combines heterogeneous operators and data patterns. It integrates dense projections, element-wise operations, and recursive accumulation, while involving both static parameters and dynamic states with distinct access patterns, making unified and efficient execution difficult. Third, recursive state propagation introduces strong dependencies that limit parallelism and result in bandwidth-dominated execution.

\begin{figure}[t]
    \centering
    \includegraphics[width=\linewidth]{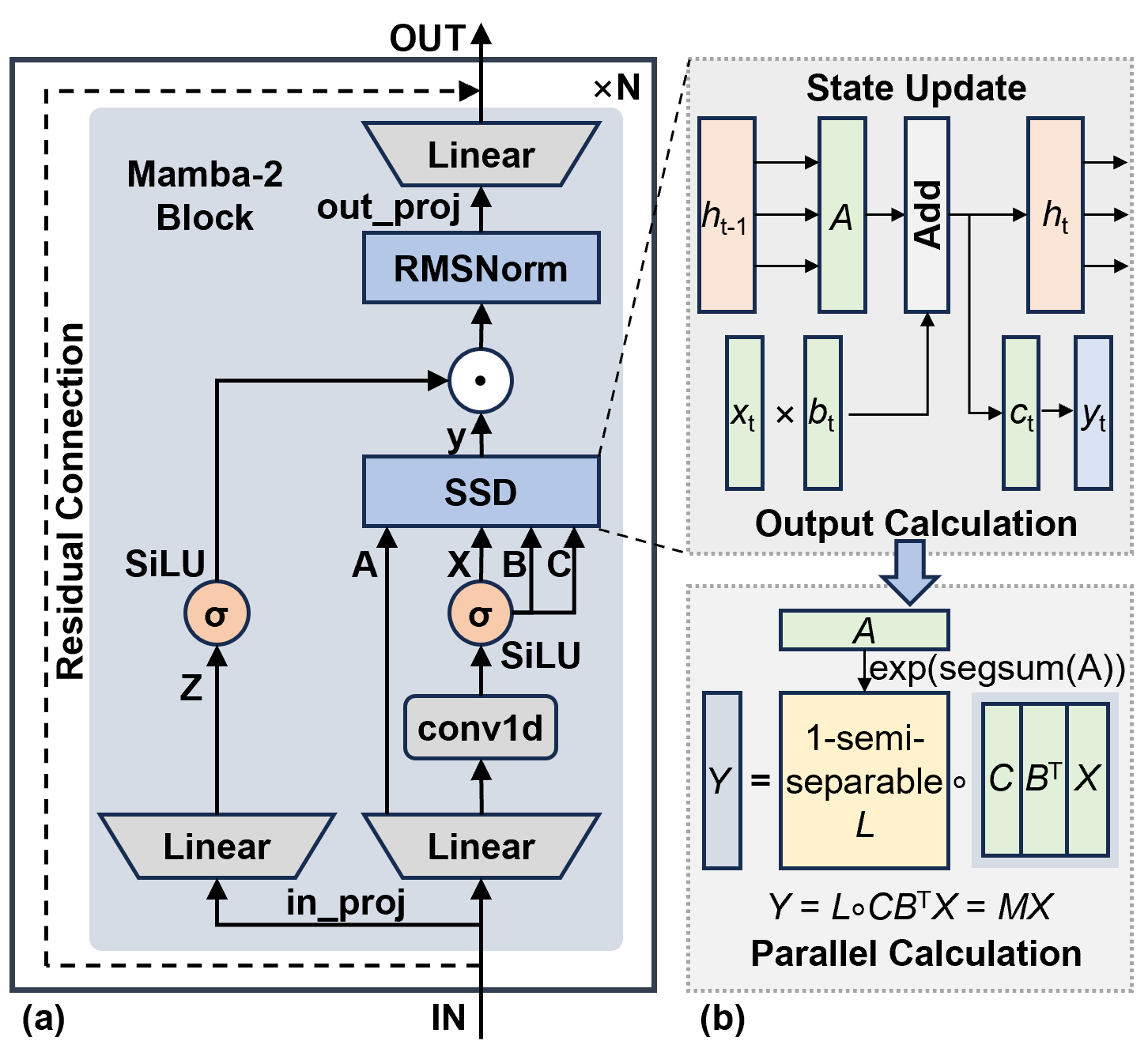}
    \caption{(a) Computational structure of the Mamba-2 block. (b) Detailed internal dataflow of the SSD block.}
    \label{fig:mamba2_flow}
\end{figure}

To address these challenges, we propose HEMERA, a heterogeneous memory-centric accelerator for efficient Mamba-2 inference under edge constraints. HEMERA adopts an algorithm--hardware co-design that reassociates the full SSD contraction into a matrix-free streaming schedule while preserving its semantics. Our main contributions are summarized as follows:

\begin{enumerate}[leftmargin=2em, itemsep=1pt, topsep=1pt]
    \item \textbf{Recursive Execution Reformulation:} We reorganize SSD into three hardware-aligned primitives---\emph{injection}, \emph{state}, and \emph{decay}---to avoid quadratic intermediate materialization and enable unified streaming execution.

    \item \textbf{Memory-Centric Co-Designed Hierarchy:} We propose a co-designed NAND--eDRAM--SRAM hierarchy with heterogeneous processing elements (PEs), mapping dense projections to digital CIM (DCIM) and recursive element-wise updates to a streaming engine while accommodating both static weights and dynamic states.

    \item \textbf{Streaming Recursive Engine (SRE):} We develop a dedicated engine for recursive state propagation, using a systolic structure to map unavoidable sequential dependencies onto streaming execution and reduce serialization overhead.
\end{enumerate}

\section{Background}
\subsection{Computation Model of Mamba-2}
Fig.~\ref{fig:mamba2_flow}(a) illustrates the computation flow of a Mamba-2 block. Linear projections generate the state transition parameter $A$, intermediate features for SSD computation, and an auxiliary branch $Z$ for gating. The intermediate features are then processed by a 1D convolution followed by a SiLU activation to produce the key SSD tensors, including the input feature map $X$ and parameter matrices $B$ and $C$.

\begin{figure}[t]
    \centering
    \includegraphics[width=\linewidth]{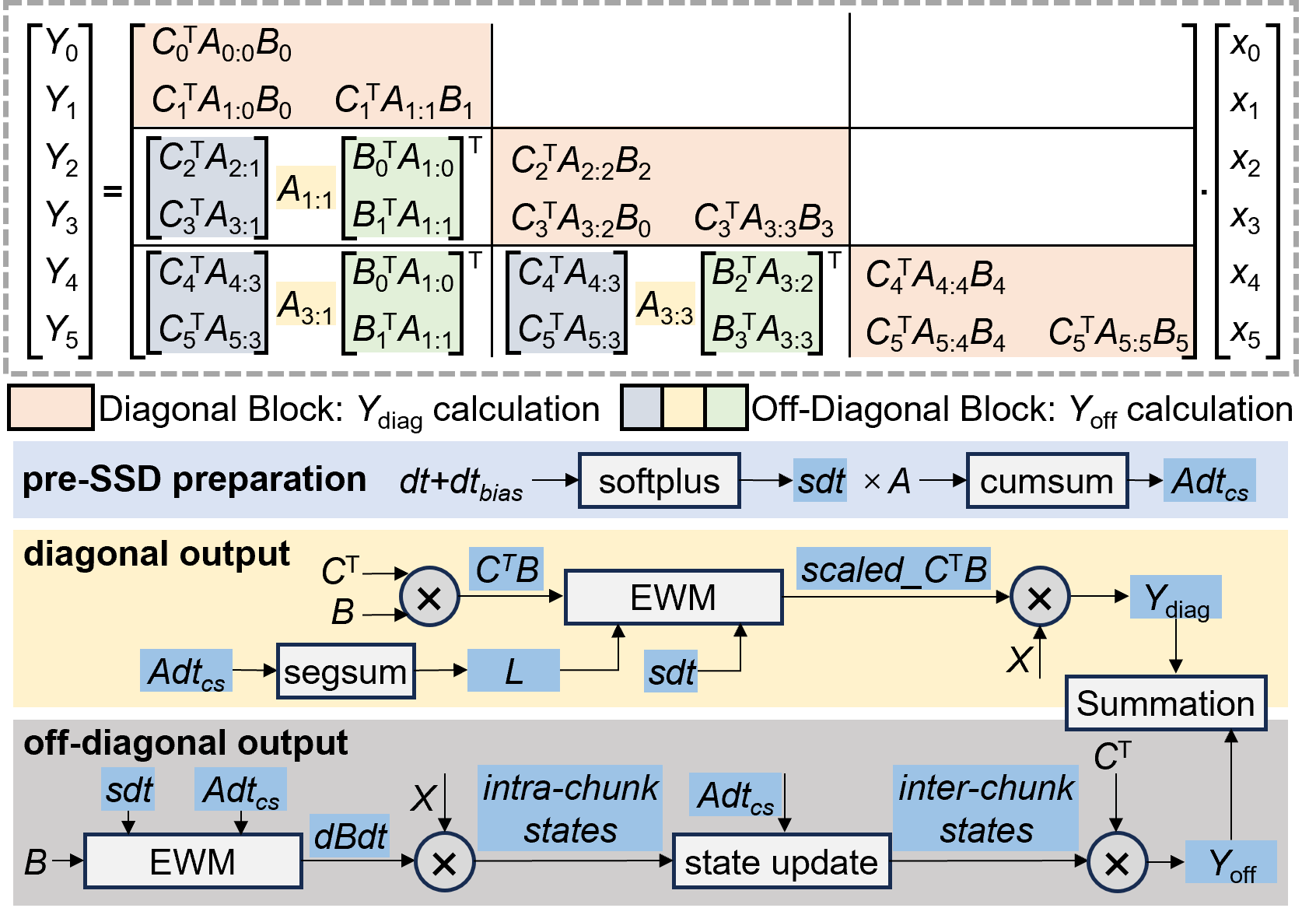}
    \caption{Chunk-wise matrix-oriented SSD dataflow with intra- and inter-chunk computation.}
    \label{fig:fig2}
\end{figure}

\input{Tables/table1}

As shown in Fig.~\ref{fig:mamba2_flow}(b), SSD is derived from a state-space formulation~\cite{gu2024mambalineartimesequencemodeling}, where the hidden state evolves according to
\begin{equation}
    h_t=A_th_{t-1}+B_tx_t
\label{equation1}
\end{equation}
\begin{equation}
    y_t=C_th_t
\label{eq:ssd2}
\end{equation}
Here, $x_t$ corresponds to the input feature $X$ at step $t$, $h_t$ is the hidden state, and $A_t$, $B_t$, and $C_t$ are the model parameters governing state transition and output generation.

\begin{figure*}[t]
    \centering
    \includegraphics[width=\textwidth]{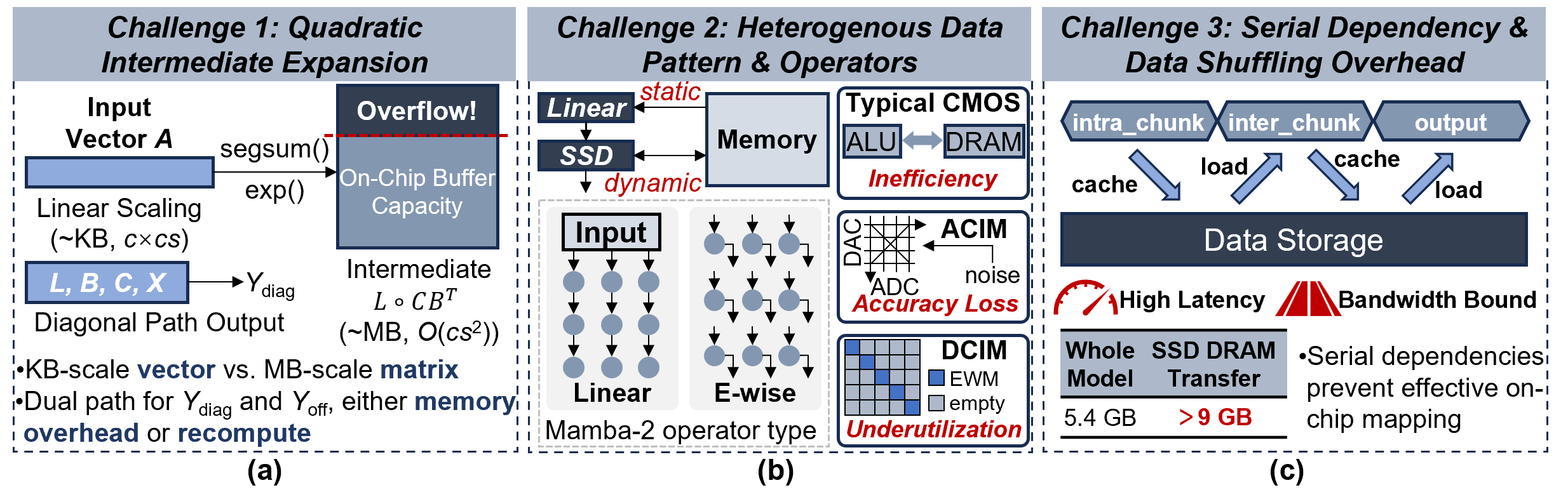}
    \caption{Core challenges for efficient Mamba-2 deployment: (a) Intermediate data explosion; (b) Operator heterogeneity requiring a unified and flexible compute paradigm; (c) High latency and bandwidth bottlenecks.}
    \label{fig:fig3}
\end{figure*}

From a matrix perspective, SSD can be equivalently expressed as
\begin{equation}
    Y=(L\circ(CB^T))X=MX
\label{eq:ssd_matrix}
\end{equation}
where $\circ$ denotes element-wise multiplication (EWM). The matrix $L=e^{\mathrm{segsum}(A)}$ is a structured lower-triangular 1-semi-separable (1SS) matrix that encodes sequence dependencies across time steps. This formulation establishes a connection between SSD and attention-like computation while exposing its structured matrix operations.

\subsection{Hardware Implication of SSD Execution}
The lower-triangular matrix $L$ encodes structured dependencies across the sequence. To expose parallelism, GPU implementations partition the input into fixed-length chunks. This strategy aligns with SSD's block structure and enables parallel intra-chunk computation while preserving cross-chunk state propagation.

For a sequence of length $s$, the sequence is partitioned into $c$ chunks of size $cs$. The dimensions of key tensors (e.g., $A$, $L$, $X$, $B$, $C$, $Y_{\mathrm{diag}}$, and $Y_{\mathrm{off}}$) are summarized in Tab.~\ref{table1}~\cite{mamba2,11159607}. Under this representation, computation is decomposed into intra-chunk operations that capture local dependencies and inter-chunk state propagation that models cross-chunk dependencies, corresponding to the block-diagonal and off-diagonal structure of $L$. As illustrated in Fig.~\ref{fig:fig2}, execution consists of pre-SSD preparation, intra-chunk output generation, and inter-chunk state propagation.

\input{Tables/table2}

In intra-chunk computation, a tensor contraction between $C$ and $B$ forms a pairwise interaction matrix, which is then scaled by accumulated decay factors to produce $\mathit{scaled\_C}^{\mathrm{T}}B$. Multiplying this tensor by $X$ yields $Y_{\mathrm{diag}}$, the output from intra-chunk interactions. In the staged matrix-oriented dataflow shown in Fig.~\ref{fig:fig2}, intermediate tensors are exposed between operators, as highlighted in blue. Among them, $\mathit{scaled\_C}^{\mathrm{T}}B$ dominates the formulation-level storage footprint when materialized, as its dimensionality scales quadratically with the chunk size $cs$.

In inter-chunk state propagation, hidden states are passed across chunks to maintain global recurrence. Within each chunk, intermediate states are generated through element-wise decay and tensor contractions and then sequentially propagated, with the final state of one chunk initializing the next. This dependency introduces partial serialization, limiting parallel efficiency and requiring synchronization across chunk boundaries. The resulting output $Y_{\mathrm{off}}$ captures long-range dependencies across chunks.

\section{Challenges and Motivation}
SSD’s dual formulation introduces a structural mismatch between algorithmic abstraction and hardware execution, arising from the need to simultaneously support recursive state evolution, intermediate materialization, and heterogeneous operator patterns. While the matrix view enables parallelism, practical inference must still accommodate these execution characteristics. As illustrated in Fig.~\ref{fig:fig3}, they give rise to three fundamental challenges that motivate the design of HEMERA.

The first challenge arises from the intermediate expansion induced by matrix-oriented SSD execution. When temporal interactions are explicitly materialized within each processing chunk, the intermediate storage complexity grows as $O(cs^2)$. While such storage overhead can be accommodated on high-end GPUs, edge-constrained accelerators operate under much tighter on-chip memory budgets. As shown in Tab.~\ref{tab:table2} and Fig.~\ref{fig:fig3}(a), for a 2.8B model, the compact 640~KB decay representation participates in forming a structured intermediate of up to 160~MB when materialized, placing substantial pressure on on-chip buffers.

The second challenge arises from the heterogeneous operators and data patterns in Mamba-2 blocks. Dense projections rely on static parameters and high-throughput accumulation, whereas recursive updates and element-wise operations involve dynamic states with fine-grained, dependency-sensitive execution. As shown in Fig.~\ref{fig:fig3}(b), these differences impose conflicting requirements on computation and data access. Analog CIM (ACIM) enables efficient accumulation but suffers from noise that degrades recursive state propagation, while DCIM preserves precision but may underutilize fine-grained operations. Consequently, no single conventional compute paradigm efficiently supports all SSD operators.

The third challenge stems from the intrinsic temporal dependency in SSD’s recursive formulation. Each hidden state depends on its predecessor, limiting independent execution across time steps. Consequently, states must be propagated sequentially and may incur repeated transfers across the memory hierarchy under conventional mappings, particularly between intra- and inter-chunk updates. As shown in Fig.~\ref{fig:fig3}(c), in the staged matrix-oriented execution profiled for a 2.8B model with $cs=256$, SSD-related DRAM traffic exceeds 9~GB~\cite{11159607}, surpassing the model-parameter footprint. This limited parallelism and frequent state transfer make SSD execution bandwidth-bound rather than compute-bound.

These challenges motivate HEMERA’s layered co-design. First, HEMERA reformulates SSD into a matrix-free recursive dataflow that avoids quadratic intermediate materialization and unifies intra- and inter-chunk updates on a shared datapath. Second, a heterogeneous fabric maps projection-dominated operations to DCIM and precision-sensitive recursion to SRE, aligning the execution paradigm with operator characteristics. Third, the unavoidable recurrence is mapped onto a dedicated streaming SRE to internalize state propagation and reduce dependency-induced data movement. Together, these design principles enable efficient large-scale Mamba-2 inference under edge constraints.

\begin{figure}[t]
    \centering
    \includegraphics[width=\linewidth]{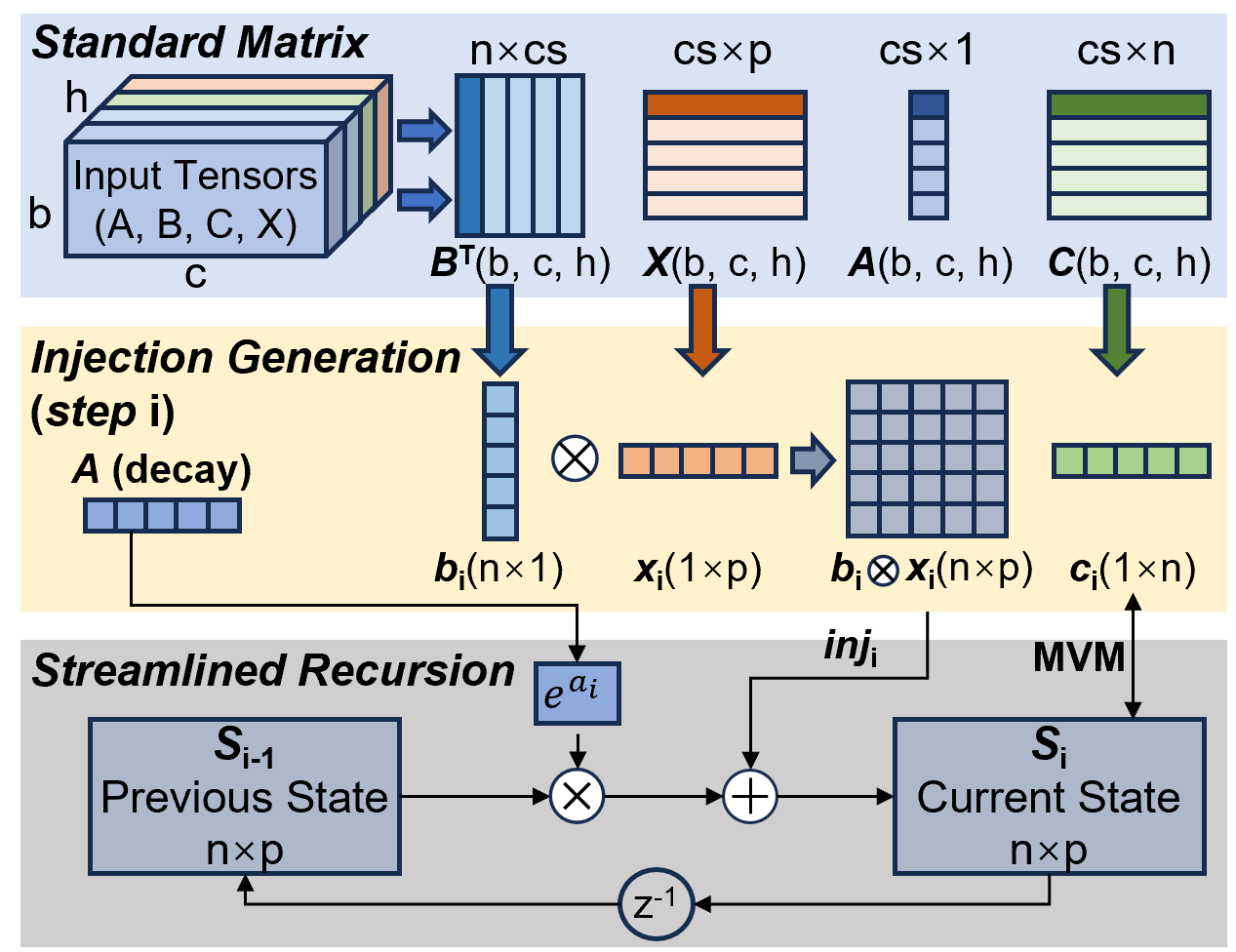}
    \caption{Recursive computation of $Y_{\mathrm{diag}}$ in the proposed SSD reformulation.}
    \label{fig:fig4}
\end{figure}

\section{SSD Reformulation and Streaming Execution}
\subsection{Algebraic Recursive Reformulation of SSD}
We begin with the algebraic structure of the 1SS matrix underlying
SSD. Let $d_t=e^{a_t}$ denote the decay at step $t$. For sequence indices $0,\ldots,T-1$, each element of
$L=\exp(\mathrm{segsum}(A))$ is

\begin{equation}
L_{t,i}=
\begin{cases}
\displaystyle\prod_{k=i+1}^{t}d_k, & 0\le i\le t,\\
0, & i>t,
\end{cases}
\label{eq:L_structure}
\end{equation}
where an empty product equals one, yielding $L_{t,t}=1$.

For clarity, batch and head indices are omitted. The full
matrix-form SSD output at step $t$ is

\begin{equation}
\begin{aligned}
Y_t
&=\sum_{i=0}^{t}L_{t,i}
    \left(C_t^{T}B_i\right)X_i \\
&=C_t^{T}\sum_{i=0}^{t}
    L_{t,i}\left(B_i\otimes X_i\right),
\end{aligned}
\label{eq:ssd_expansion}
\end{equation}
where $\otimes$ denotes the outer product. Following the notation
in Fig.~\ref{fig:fig4}, we define the token-wise injection
$\mathrm{inj}_t=B_t\otimes X_t$ and the recursive state $S_t$ as

\begin{equation}
\begin{aligned}
\mathrm{inj}_t &= B_t\otimes X_t,\\
S_t &=
\begin{cases}
\mathrm{inj}_t, & t=0,\\
d_tS_{t-1}+\mathrm{inj}_t, & t>0,
\end{cases}\\
Y_t &= C_t^{T}S_t .
\end{aligned}
\label{eq:recurrence}
\end{equation}

Expanding the recurrence gives
$S_t=\sum_{i=0}^{t}L_{t,i}\mathrm{inj}_i$; substituting it into
the readout $C_t^{T}S_t$ exactly recovers
Eq.~\ref{eq:ssd_expansion}. Therefore, the proposed
injection--state--decay schedule is algebraically equivalent to the
original SSD computation, without pruning or approximation. It
evaluates the same contraction without explicitly constructing $L$
or the quadratic pairwise-interaction tensor, exposing a
streaming-friendly dataflow for hardware execution.

\subsection{Recursive SSD Calculation}
Based on Eq.~\ref{eq:recurrence}, HEMERA evaluates SSD through token-wise injection, state propagation, and output readout. At each step, $\mathrm{inj}_t=B_t\otimes X_t$ is generated locally, accumulated into $S_t$ through the decay-modulated update, and projected by $C_t^T$ to produce $Y_t$. As shown in Fig.~\ref{fig:fig4}, this execution order avoids explicitly constructing the quadratic pairwise tensor $\mathit{scaled\_C}^{\mathrm{T}}B$ and exposes a streaming dataflow.

For chunked SSD, the same affine update primitive supports both intra- and inter-chunk contributions. Intra-chunk processing accumulates local injections from the chunk boundary, whereas inter-chunk processing seeds the recurrence with the terminal state propagated from the preceding chunk. Thus, the two paths share the same computation logic while differing in state initialization and decay scope. The original temporal dependency is preserved, but state propagation is internalized within a streaming datapath.

\input{Tables/table3}

Compared with matrix-oriented SSD execution, the recursive schedule reduces intermediate storage from $O(cs^2)$ to $O(cs\times p+n\times p)$, comprising a $cs\times p$ output buffer and one $n\times p$ state. Tab.~\ref{tab:table3} reports formulation-level operation counts for the matrix-oriented $Y_{\mathrm{diag}}$ evaluation rather than runtime speedup over a fused GPU kernel. For a fixed sequence length, HEMERA avoids the chunk-size-dependent pairwise contraction, leaving token-wise outer product, state update, and output readout.

\begin{figure}[t]
    \centering
    \includegraphics[width=\linewidth]{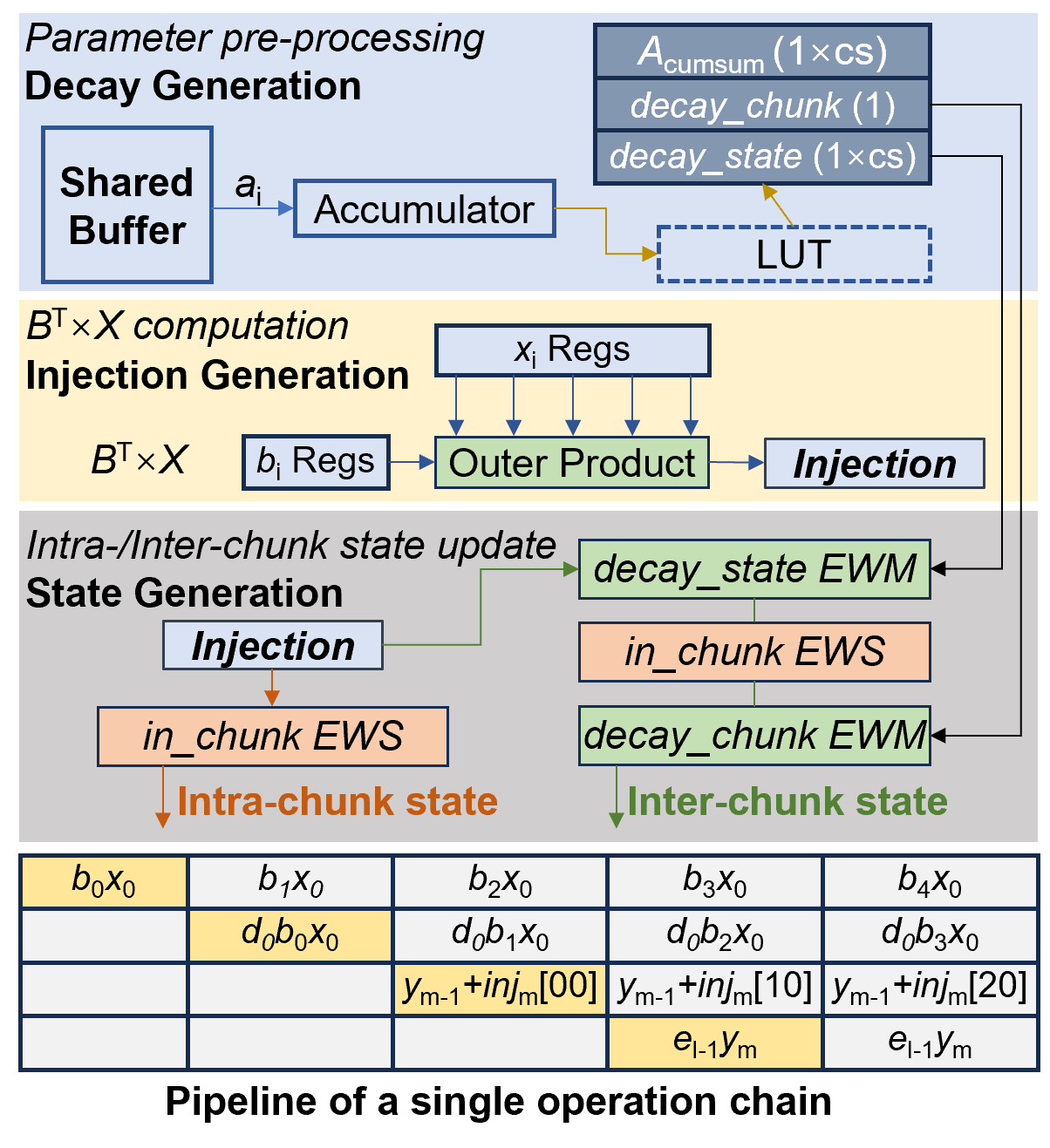}
    \caption{Workflow and pipeline for SSD recursion mapping.}
    \label{fig:fig5}
\end{figure}

\section{Hardware Design Details}
\subsection{Recursive Execution Logic and Dataflow Mapping}
To map the matrix-free SSD recurrence onto hardware, HEMERA
organizes Eq.~\ref{eq:recurrence} into the three-stage streaming
flow shown in Fig.~\ref{fig:fig5}: decay generation, injection
generation, and state propagation. The flow distinguishes local
injection generation from local and cross-SRE state propagation,
avoiding explicit construction of the quadratic pairwise tensor.

\textbf{Stage 1: Decay Generation.}
Decay-related parameters are loaded from the shared buffer and
accumulated by the Prefix Sum Unit (PSU). The resulting prefix
values are converted through an exponential look-up table (LUT)
into decay terms $d_t$ for intra-chunk updates and inter-chunk state
propagation.

\textbf{Stage 2: Injection Generation.}
The operands $B_t$ and $X_t$ are streamed through local registers
and an outer-product unit to generate the token-wise injection
$\mathrm{inj}_t=B_t\otimes X_t$. Each injection is forwarded
directly to the state-update datapath without materializing pairwise
interaction tensors.

\textbf{Stage 3: State Propagation.}
The unified datapath performs
$S_t=d_tS_{t-1}+\mathrm{inj}_t$. For intra-chunk computation,
$S_{t-1}$ is retrieved from the local state buffer. For inter-chunk
propagation, the state is received from the upstream SRE and
forwarded downstream after the update. Both modes reuse the same
multiply--accumulate logic while differing in the source of the
previous state and the decay scope. The updated state is subsequently
contracted with $C_t$ to produce $Y_t=C_t^TS_t$. This mapping
preserves the temporal dependency while internalizing state movement
within the streaming pipeline. 

\begin{figure*}[t]
    \centering
    \includegraphics[width=\textwidth]{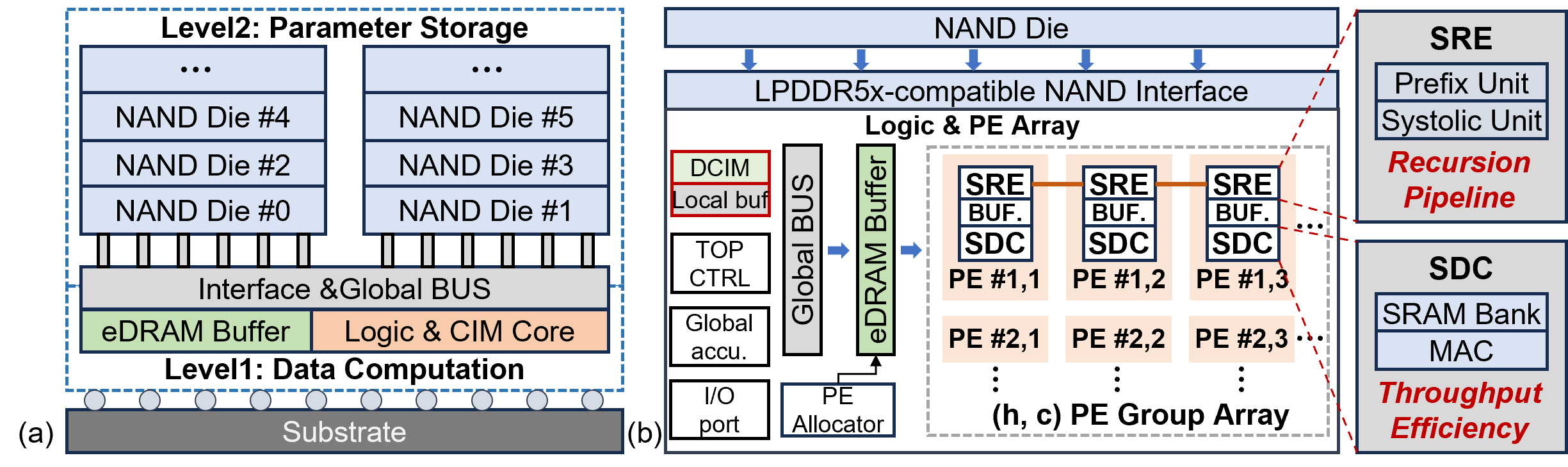}
    \caption{HEMERA hardware architecture.}
    \label{fig:fig6}
\end{figure*}

\subsection{Overall Architecture}
Building on the recursive execution and dataflow mapping, Mamba-2
computation can be divided into two complementary categories. The
first consists of dense linear projections outside the SSD core,
dominated by matrix--vector and matrix--matrix multiplications with
abundant parallelism. The second comprises recursive state updates
and element-wise operations within SSD, whose temporal dependencies
require continuous state propagation. Therefore, an efficient
architecture must jointly support high-throughput dense computation
and low-latency streaming execution.

As shown in Fig.~\ref{fig:fig6}, HEMERA adopts a two-level physical
organization that implements a three-tier NAND--eDRAM--SRAM memory
hierarchy. Level 2 consists of 3D-stacked NAND dies for high-density
parameter storage, with weights streamed layer by layer through an
LPDDR5x-compatible interface~\cite{10946816}. Level 1 contains the
compute tier, where an on-chip eDRAM buffer decouples NAND accesses
from computation and distributes staged data through a global
interconnect to the compute fabric.

The compute fabric comprises a PE array orchestrated by a top-level
controller and PE allocator, together with a global accumulator for
cross-PE reduction and an I/O port for external communication. Each
PE integrates an SRE and an SRAM-based digital CIM (SDC) core. The
SRE performs decay generation and recursive state propagation,
whereas the SDC core executes dense projections using parallel SRAM
banks and MAC units. Co-locating the two engines enables direct data
exchange between projection and recursion paths, reducing
intermediate buffering and movement across the memory hierarchy.

\subsection{PE Microarchitecture}

\begin{figure}[t]
    \centering
    \includegraphics[width=\linewidth]{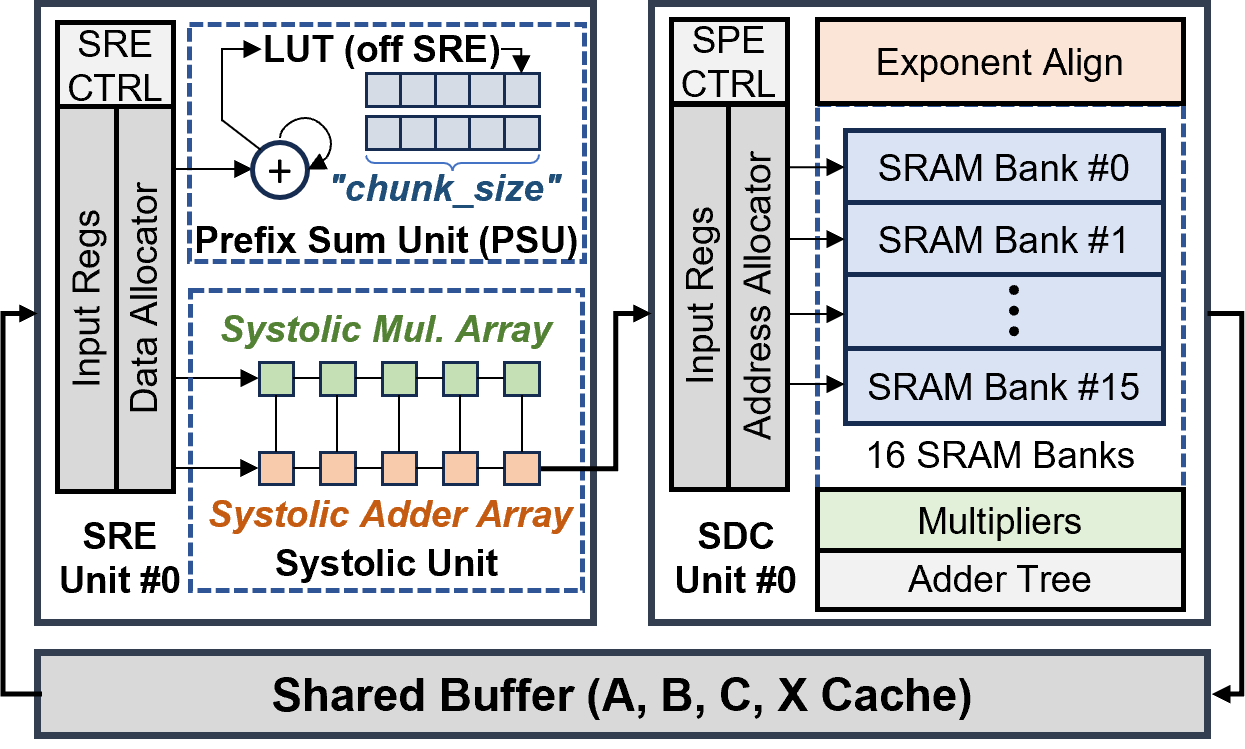}
    \caption{Detailed architecture of a single PE.}
    \label{fig:fig7}
\end{figure}

As shown in Fig.~\ref{fig:fig7}, each PE integrates a shared buffer,
an SRE, and an SDC core. For dense projections, weights streamed
layer by layer from Level~2 NAND are buffered locally and supplied
to the SDC core. During SSD execution, the shared buffer stores the
dynamic tensors ($A$, $B$, $C$, and $X$) and active states used by
both engines, enabling local reuse and direct data exchange between
projection and recursive operations.

The SRE implements the matrix-free recurrence using two coupled
sub-units: a Prefix Sum Unit (PSU) and a Systolic Unit (SU). The PSU accumulates the $a_t$ values and generates the corresponding
decay factors $d_t$ through an exponential look-up table (LUT).
The SU receives token-wise injections $\mathrm{inj}_t$ and performs
$S_t=d_tS_{t-1}+\mathrm{inj}_t$. It accesses locally buffered states for
intra-chunk execution and uses pipelined links for inter-chunk state
propagation, avoiding repeated state transfers through the global
memory hierarchy. Because the SRE implements the affine update
$S_t=d_tS_{t-1}+\mathrm{inj}_t$, it can also support
Mamba-1-style selective scans through state-dimension
reconfiguration.

The SDC core performs dense projections inside and outside the SSD
block. It comprises 16 parallel SRAM banks, a shared MAC array, and
exponent-alignment logic for floating-point computation. Weights
and activations are distributed across the banks to provide parallel
access, while the MAC array and exponent-alignment logic perform
numerically consistent accumulation.

\begin{figure}[t]
    \centering
    \includegraphics[width=\linewidth]{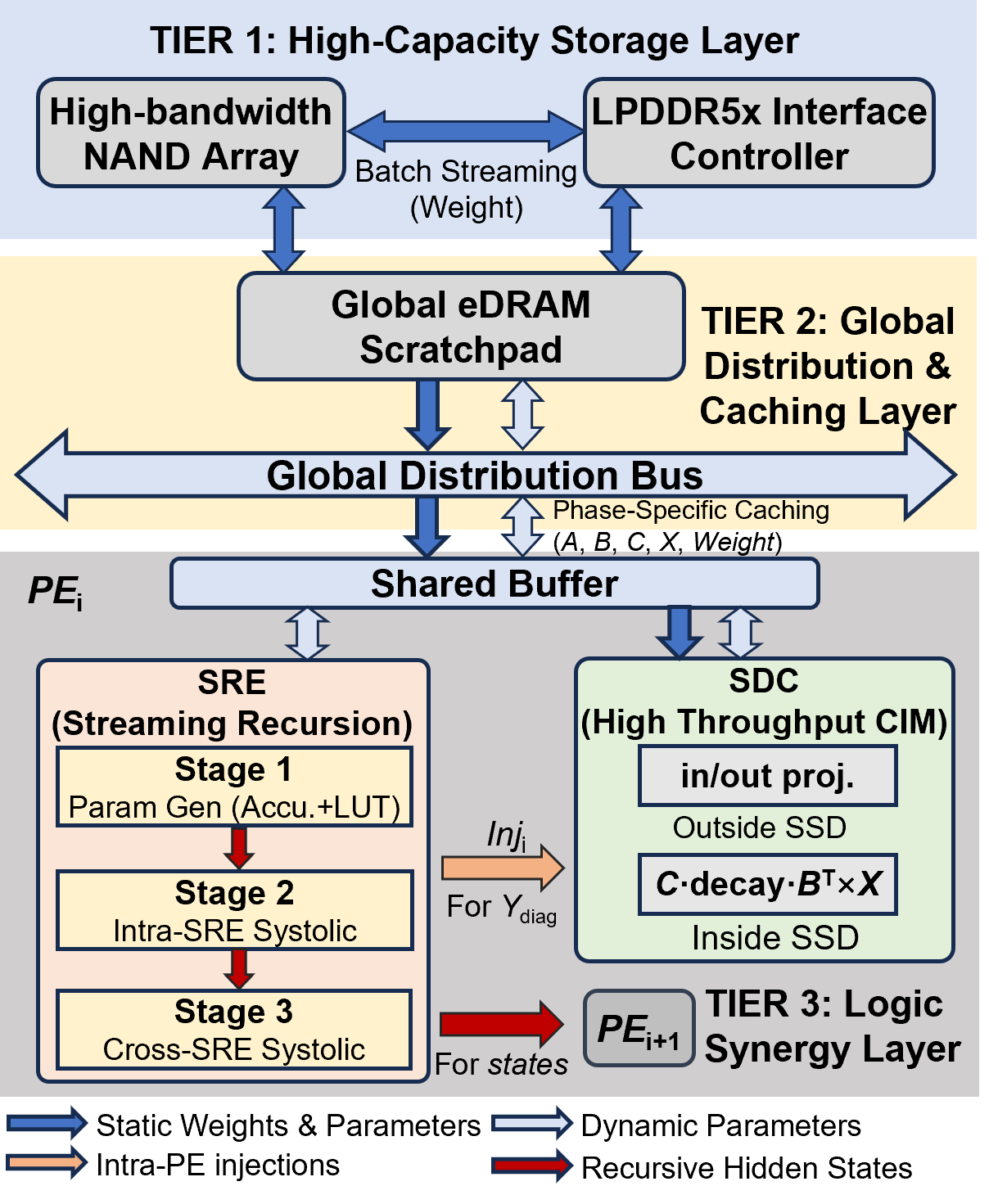}
    \caption{Tiered dataflow architecture of HEMERA for Mamba-2 inference.}
    \label{fig:fig8}
\end{figure}

\subsection{End-to-End Execution Mapping}
Fig.~\ref{fig:fig8} illustrates HEMERA's end-to-end mapping across
the NAND, eDRAM, and PE-local SRAM tiers. NAND streams layer-wise
weights, eDRAM stages and distributes data, and the PE-local SRAM
feeds the SDC and SRE engines. Dense projections are mapped to the
SDC cores, while the SRE executes
$S_t=d_tS_{t-1}+\mathrm{inj}_t$. Intra-chunk states are retrieved
locally, whereas inter-chunk states are forwarded through neighboring
PEs, localizing recursive propagation and reducing global data
movement.

\begin{figure*}[t]
    \centering
    \includegraphics[width=\textwidth]{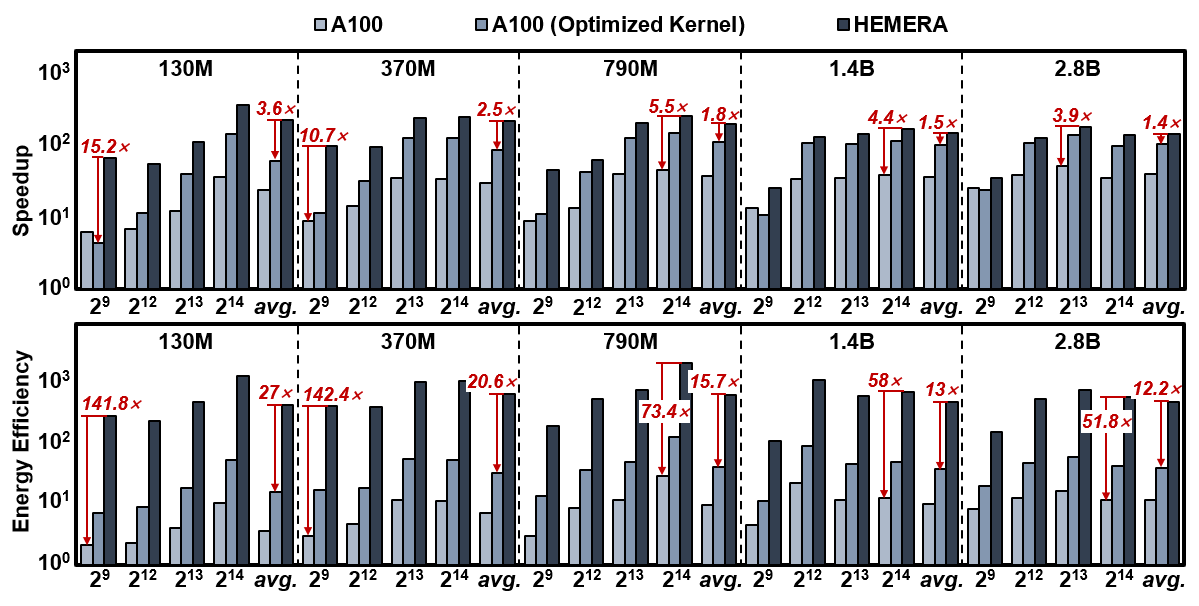}
    \caption{Performance comparison between HEMERA and the NVIDIA A100 GPU relative to the CPU baseline.}
    \label{fig:fig9}
\end{figure*}

\begin{figure}[t]
    \centering
    \includegraphics[width=\linewidth]{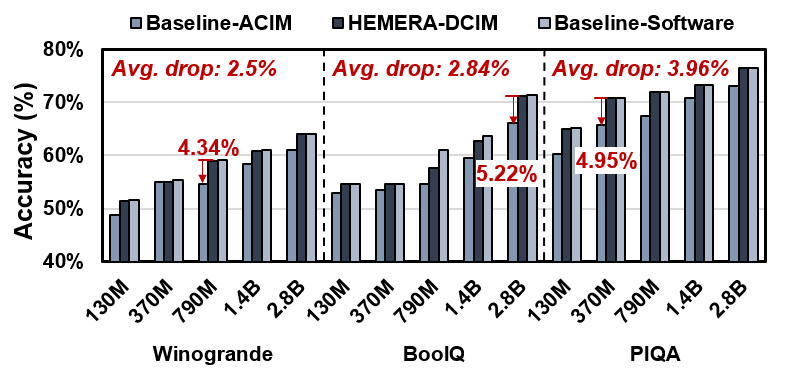}
    \caption{End-to-end accuracy comparison among the software baseline, HEMERA's DCIM execution, and the ACIM alternative.}
    \label{fig:fig10}
\end{figure}

\input{Tables/table4}
\input{Tables/table5}

\section{Evaluation}
\subsection{Experiment Setup}

\textbf{Methodology.}
We evaluate HEMERA on Mamba-2 models ranging from 130M to
2.8B parameters. Unless otherwise specified, $chunk\_size$, $d\_state$, and $d\_head$ are set to 64. The compute fabric is implemented in Verilog HDL
and synthesized in a 22-nm CMOS process at 0.9~V, with timing,
area, and power extracted from post-synthesis results. On-chip
memories are modeled using CACTI-7~\cite{10.1145/3085572}, while
the 3D-stacked NAND subsystem is characterized using vendor-reported
parameters~\cite{kioxia_fl6_2022,9063154}. End-to-end performance
is obtained through a trace-driven framework that combines
RTL-measured compute latency with an analytical memory model
derived from Mamba-2 operator traces.

\textbf{Baseline.}
HEMERA is compared against an Intel Xeon Platinum CPU and an
NVIDIA A100 GPU~\cite{nvidia_a100_2024} across sequence lengths
from 128 to 16384 and model sizes from 130M to 2.8B. A
PyTorch/CUDA FP16/BF16 implementation is retained for
operator-level SSD profiling because it exposes intra-chunk
computation, inter-chunk state propagation, and intermediate-data
movement. The official optimized fused Mamba-2 kernel is used as
the end-to-end A100 baseline under the same model, sequence-length,
and precision settings.   A100 energy is calculated using profiled runtime power. 
We also provide a contextual comparison
with representative Mamba accelerators.

\textbf{HEMERA Configuration.}
Tab.~\ref{tab:table4} summarizes the hardware configuration, and
Tab.~\ref{tab:table5} reports the system-level area/footprint and
power accounting. HEMERA operates at 1~GHz and supports BF16
execution in the SDC cores. The storage subsystem comprises 16 NAND
dies connected through an 8-channel LPDDR5x-compatible interface
with 136.5~GB/s peak bandwidth
\cite{micron_lpddr5_datasheet,morra_infineon_lpddr_2023}.
Weights are streamed sequentially at layer granularity and staged in
a 32~MB eDRAM scratchpad. On-chip SRAM totals 8.6~MB, including
1.92~MB for the SDC cores and 6.7~MB for shared buffers. The
$(24,10)$ configuration denotes 24 head-parallel lanes and 10
chunk-parallel slots, totaling 240 PEs. With a chunk size of 64, one
execution wave covers up to 640 tokens per mapped head group;
additional heads and longer sequences are processed through repeated
waves. The PE allocator schedules parallelism across heads and chunks.

\subsection{End-to-End Performance}

\textbf{Latency Speedup.}
As shown in Fig.~\ref{fig:fig9}, HEMERA achieves up to
$15.2\times$ speedup over the PyTorch/CUDA A100 implementation
and $1.4\times$--$3.6\times$ average speedup over the official
optimized fused kernel. GPU-side fusion narrows the gap, while
HEMERA retains its advantage by avoiding quadratic intermediate
construction and localizing recursive state propagation. 

\textbf{Energy Efficiency.}
HEMERA achieves $12.2\times$--$27.0\times$ higher energy
efficiency than the optimized fused-kernel A100 by reducing
off-chip traffic and using SRAM-based digital CIM.  For the 2.8B
model at a sequence length of 4096, HEMERA requires 130.00~ms,
corresponding to 31.738~$\mu$s/token and 0.9541~mJ/token.

\begin{figure}[t]
    \centering
    \includegraphics[width=\linewidth]{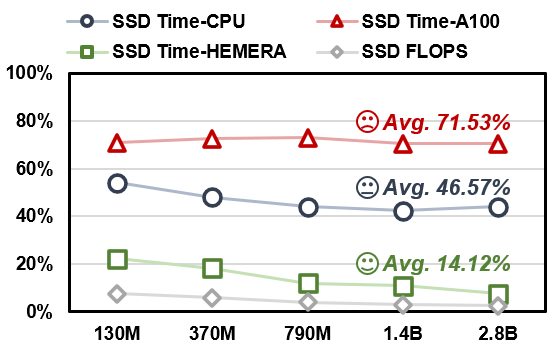}
    \caption{Time ratio of SSD computation in end-to-end Mamba-2 execution across model scales.}
    \label{fig:fig11}
\end{figure}

\textbf{Accuracy Robustness.}
Fig.~\ref{fig:fig10} compares the PyTorch software baseline,
HEMERA's DCIM execution, and an ACIM alternative. HEMERA closely
tracks the software baseline under digital finite-precision execution,
whereas ACIM incurs average accuracy drops of $2.5\%$--$3.96\%$
and a maximum drop of $5.22\%$ under characterized MAC-level
non-idealities~\cite{10031424,9441013,11044213}.

\subsection{Execution Breakdown and Utilization}
To understand the sources of performance gains, we analyze the
runtime breakdown of the SSD core together with the average compute
utilization during Mamba-2 inference.

\textbf{SSD Time Breakdown.}
Fig.~\ref{fig:fig11} compares the SSD time ratio across CPU, A100, HEMERA. The
original A100 implementation is retained here because it exposes
individual SSD stages for operation-level profiling, whereas the
official fused kernel combines these stages. Despite accounting for
only $3\%$--$7.5\%$ of total FLOPs, SSD occupies $71.53\%$ and
$46.57\%$ of end-to-end latency on A100 and CPU, respectively. In
contrast, HEMERA reduces this ratio to $14.12\%$ on average by
avoiding explicit $\mathit{scaled\_C}^{\mathrm{T}}B$ construction
and internalizing state propagation within the streaming-recursive
pipeline.

\begin{figure}[t]
    \centering
    \includegraphics[width=\linewidth]{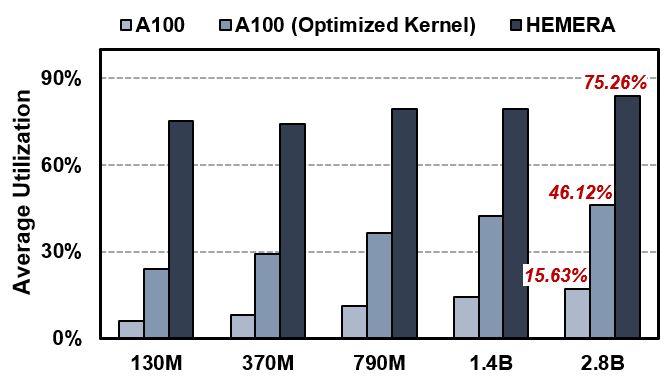}
    \caption{Average compute utilization of the original A100
    implementation, the official optimized fused kernel, and HEMERA
    across model scales.}
    \label{fig:fig12}
\end{figure}

\input{Tables/table6}

\textbf{Hardware Utilization.}
Fig.~\ref{fig:fig12} compares average compute utilization over
sequence lengths from 128 to 16384. The optimized kernel
substantially improves A100 utilization by reducing operator and
memory overheads. Nevertheless, HEMERA consistently achieves
higher utilization across all model scales. For the 2.8B model, the
utilization increases from $15.63\%$ on the original A100
implementation to $46.12\%$ with the optimized kernel, while HEMERA
reaches $75.26\%$, corresponding to $4.8\times$ and $1.63\times$
higher utilization, respectively. This remaining advantage results
from the specialized execution of dense projections and localized
recursive state propagation, which reduces data-movement overhead
and dependency-induced stalls.

\subsection{State-of-the-Art Comparison}
Tab.~\ref{tab:table6} provides a contextual comparison between
HEMERA and representative Mamba/SSM accelerators. These works
differ in target model, platform, precision, and workload; therefore,
the normalized metrics should not be interpreted as a direct ranking.
Prior designs primarily focus on operator-level optimization,
reconfigurable dataflow, or speculative execution. In contrast,
HEMERA combines a tiered NAND--eDRAM--SRAM hierarchy with a
matrix-free SSD schedule mapped to a dedicated SRE. By coupling
digital-CIM-based dense projection with localized recursive
execution, HEMERA supports efficient end-to-end Mamba-2 inference
under edge constraints.

\section{Conclusion}
This paper presents HEMERA, a heterogeneous memory-centric
accelerator for efficient Mamba-2 inference. By reassociating the
full SSD contraction into an algebraically equivalent matrix-free
streaming recurrence and mapping it onto a dedicated SRE tightly
coupled with digital-CIM projection cores, HEMERA reduces
quadratic intermediate storage while preserving the temporal
dependency of state evolution. For the 2.8B Mamba-2 model,
HEMERA achieves a $1.4\times$ average latency speedup and a
$12.2\times$ energy-efficiency improvement over the official
optimized fused-kernel NVIDIA A100 baseline, while reducing the
SSD execution-time ratio to below $15\%$. Future work will explore
extending this recursion-centric design framework to broader
state-space and non-attention architectures under edge constraints.

\section{Acknowledgments}
This work was supported by the National Natural Science Foundation of China (62341407, 62322401 and 62406008), Beijing Natural Science Foundation (L223004), and in part by the "111" Project (B18001).

\bibliographystyle{ACM-Reference-Format}
\bibliography{refs}

\end{document}

%% file: Tables/table1.tex
\renewcommand{\arraystretch}{0.9}
\setlength{\tabcolsep}{4pt}
\begin{table}[t]
\vspace{1pt}
\centering
\caption{Dimension index and semantic meanings of key tensors in SSD.}
\vspace{1pt}
\label{table1}

\renewcommand{\tabularxcolumn}[1]{m{#1}}

\begin{tabularx}{\columnwidth}{@{}
    >{\centering\arraybackslash}X
    >{\centering\arraybackslash}X
    >{\centering\arraybackslash}X@{}}
\toprule
\textbf{Symbol} & \textbf{Dimension Index} & \textbf{Semantic Meaning} \\
\midrule
$A$ & $(b, h, c, cs)$ & Decay vector \\
$L$ & $(b, h, c, cs, cs)$ & 1-semi-separable matrix \\
$X$ & $(b, c, cs, h, p)$ & Input projections \\
$B, C$ & $(b, c, cs, h, n)$ & Evolution parameters \\
$Y_{\mathrm{diag}}$ & $(b, c, cs, h, p)$ & Diagonal output \\
$Y_{\mathrm{off}}$ & $(b, c, cs, h, p)$ & Off-diagonal output \\
\bottomrule
\end{tabularx}
\vspace{1pt}
\end{table}

%% file: Tables/table2.tex
\renewcommand{\arraystretch}{0.9}
\setlength{\tabcolsep}{1pt}
\begin{table}[t]
\vspace{1pt}
\centering
\caption{Memory footprint of $A$ and $M$ under varying model sizes (seqlen = 4096, headdim = 64).}
\vspace{1pt}
\label{tab:table2}

\renewcommand{\tabularxcolumn}[1]{m{#1}}

\begin{tabularx}{\columnwidth}{@{}
>{\centering\arraybackslash}X
>{\centering\arraybackslash}X
>{\centering\arraybackslash}X
>{\centering\arraybackslash}X@{}}
\toprule
\textbf{Model dimension} & \textbf{$A$} & \textbf{$cs = 64$} & \textbf{$cs = 256$} \\
\midrule
768 (130M)  & 192~KB & 12~MB & 48~MB \\
1024 (370M) & 256~KB & 16~MB & 64~MB \\
1536 (790M) & 384~KB & 24~MB & 96~MB \\
2048 (1.4B) & 512~KB & 32~MB & 128~MB \\
2560 (2.8B) & 640~KB & 40~MB & 160~MB \\
\bottomrule
\end{tabularx}
\vspace{1pt}
\end{table}

%% file: Tables/table3.tex
\renewcommand{\arraystretch}{0.9}
\setlength{\tabcolsep}{1pt}
\begin{table}[t]
\vspace{1pt}
\centering
\begin{threeparttable}
\caption{$Y_{\mathrm{diag}}$ operation count comparison between matrix-oriented and recursive
evaluation.}
\vspace{1pt}
\label{tab:table3}

\renewcommand{\tabularxcolumn}[1]{m{#1}}

\begin{tabularx}{\columnwidth}{@{}
>{\centering\arraybackslash}X 
>{\centering\arraybackslash}X 
>{\centering\arraybackslash}X 
>{\centering\arraybackslash}X@{}}
\toprule
\textbf{Chunk size} & \textbf{Baseline (GOPS)} & \textbf{Recursive (GOPS)} & \textbf{Reduction} \\
\midrule
64  & 68.72  & 1.6271 & 42.2$\times$ \\
128 & 137.44 & 1.6273 & 84.5$\times$ \\
256 & 274.88 & 1.6273 & 168.9$\times$ \\
512 & 549.76 & 1.6274 & 337.8$\times$ \\
\bottomrule
\end{tabularx}
\vspace{1pt}

\begin{tablenotes}[flushleft]
\footnotesize
\item[$^*$] 1 OP = 1 MAC.
\end{tablenotes}
\vspace{0pt}
\end{threeparttable}
\end{table}

%% file: Tables/table4.tex
\renewcommand{\arraystretch}{0.9}
\setlength{\tabcolsep}{4pt}
\begin{table}[t]
\vspace{1pt}
\centering
\caption{Hardware configuration of HEMERA.}
\vspace{1pt}
\label{tab:table4}

\renewcommand{\tabularxcolumn}[1]{m{#1}}

\begin{tabularx}{\columnwidth}{@{}
>{\centering\arraybackslash}p{0.38\columnwidth}
>{\raggedright\arraybackslash}X@{}}
\toprule

\multicolumn{2}{c}{\textbf{Baseline: Intel Xeon Platinum CPU}} \\
\midrule
Base Frequency & 2.7~GHz \\
Total Cores & 48 \\
Bandwidth & 20~GT/s (cache) \& 5600~MT/s (memory) \\
TDP & 350~W \\

\midrule
\multicolumn{2}{c}{\textbf{Baseline: NVIDIA A100}} \\
\midrule
Frequency & 1.41~GHz \\
Peak Performance & 312~TFLOPS (BF16) \\
Bandwidth & 2039~GB/s \\
TDP & 400~W \\

\midrule
\multicolumn{2}{c}{\textbf{Proposed: HEMERA}} \\
\midrule
Frequency & 1~GHz \\
On-chip SRAM & 8.6~MB (1.92~MB SDC + 6.7~MB shared buffer) \\
eDRAM Capacity & 32~MB (global buffer) \\
NAND & 16 dies, 8-channel LPDDR5x-compatible, 136.5~GB/s \\
Peak Performance & 246~TFLOPS (BF16 DCIM) \\
PE Configuration & $(24, 10)$ \\

\bottomrule
\end{tabularx}
\vspace{1pt}
\end{table}

%% file: Tables/table5.tex
\renewcommand{\arraystretch}{0.9}
\setlength{\tabcolsep}{1pt}

\begin{table}[t]
\centering
\caption{System-level area/footprint and power accounting of HEMERA.}
\label{tab:table5}

\begin{threeparttable}
\begin{tabularx}{\columnwidth}{@{}
>{\centering\arraybackslash}p{0.49\columnwidth}
>{\centering\arraybackslash}p{0.29\columnwidth}
>{\centering\arraybackslash}X@{}}
\toprule
\textbf{Component} &
\textbf{\shortstack{Area(mm$^2$)}} &
\textbf{Power (W)} \\
\midrule

\multicolumn{3}{c}{\textbf{Level 1: HEMERA Compute Die}} \\
\midrule
DCIM Array                 & 2.04   & 0.604 \\
Shared Buffer              & 7.36   & 2.108 \\
Intra-SDC MAC              & 28.63  & 2.703 \\
SRE                        & 10.74  & 1.014 \\
eDRAM Scratchpad           & 34.09  & 10.06 \\

\cmidrule(lr){1-3}
\textbf{Level-1 Subtotal}
  & \textbf{113.96}
  & \textbf{18.02} \\

\midrule
\multicolumn{3}{c}{%
\textbf{Level 2: Parameter Storage and Interface Dies}} \\
\midrule
NAND Array Tier
  & 145.98
  & 3.27 \\

ECC + Staging Buffers
  & 82.23
  & 2.98 \\

\begin{tabular}[c]{@{}c@{}}
LPDDR5X-Compatible Read\\
Path
\end{tabular}
  & N/A\tnote{a}
  & 5.79 \\

\midrule
\textbf{Total Architecture}
  & \textbf{145.98}
  & \textbf{30.06} \\
\bottomrule
\end{tabularx}

\begin{tablenotes}[flushleft]
\small
\item[a] The interface area is not independently modeled.
\end{tablenotes}
\end{threeparttable}
\end{table}

%% file: Tables/table6.tex
\renewcommand{\arraystretch}{0.9}
\setlength{\tabcolsep}{1pt}

\begin{table}[t]
\centering
\begin{threeparttable}

\caption{Comparison between related works.}
\label{tab:table6}

\renewcommand{\tabularxcolumn}[1]{m{#1}}

\scriptsize

\begin{tabularx}{\columnwidth}{@{}
>{\centering\arraybackslash}p{0.2\columnwidth}
>{\centering\arraybackslash}X
>{\centering\arraybackslash}p{0.16\columnwidth}
>{\centering\arraybackslash}X
>{\centering\arraybackslash}X
>{\centering\arraybackslash}X@{}}
\toprule
\textbf{Reference}
& MARCA~\cite{10.1145/3676536.3676798}
& MambaOPU~\cite{11132895}
& Mamba-X~\cite{11240777}
& SpecMamba~\cite{11240945}
& \textbf{HEMERA} \\

\midrule
Model Scope
& Mamba
& Mamba-2
& Vision Mamba
& Mamba
& Mamba-2 \\

Memory Hierarchy
& HBM+SRAM
& BRAM
& HBM
& HBM+BRAM
& \textbf{NAND--eDRAM--SRAM} \\

Precision
& 32-bit fixed-point
& FP16
& INT8
& INT4
& \textbf{BF16} \\

Optimization Focus
& Operator fusion
& Reconfigurable dataflow
& Systolic arrays
& Speculative decoding
& \textbf{Streaming recurrence} \\

Speedup$^a$
& $1.38\times$
& $<1\times$
& $2.3\times$
& $2.27\times$
& \textbf{$1.4\times$} \\

Efficiency$^a$
& $15.86\times$
& $7.46\times$
& $11.5\times$
& $5.41\times$
& \textbf{$12.2\times$} \\

SSD Breakdown
& N/A
& N/A
& 25\%--40\%$^b$
& N/A
& \textbf{$9.42\%$}$^c$ \\

\bottomrule
\end{tabularx}

\begin{tablenotes}[flushleft]
\footnotesize
\item[a] Reported relative to each work's own NVIDIA A100 baseline.
\item[b] Estimated from Fig.~17 in Mamba-X~\cite{11240777}.
\item[c] Simulated on the 2.8B model at a sequence length of 4096.
\end{tablenotes}

\vspace{-2pt}
\end{threeparttable}
\end{table}